\documentclass{article}
\usepackage{spconf,amsmath,graphicx,amssymb,amsthm,comment,bm,cite,url}
\usepackage{array}
\allowdisplaybreaks

\newcommand{\refeq}[1]{(\ref{eq:#1})}
\newcommand{\refeqs}[2]{(\ref{eq:#1}) and (\ref{eq:#2})}

\newcommand{\refsec}[1]{Section \ref{sec:#1}}
\newcommand{\refsubsec}[1]{\ref{subsec:#1}}
\newcommand{\reffig}[1]{Fig. \ref{fig:#1}}

\def\Vec#1{\boldsymbol{\mathbf{#1}}}

\def\0{{\mathbf 0}}

\def\W{{\mathbf W}}

\def\V{{\mathbf V}}

\makeatletter
\newcommand{\thickhline}{%
    \noalign {\ifnum 0=`}\fi \hrule height 1pt
    \futurelet \reserved@a \@xhline
}
\newcolumntype{"}{@{\hskip\tabcolsep\vrule width 1pt\hskip\tabcolsep}}
\makeatother

\title{StarGAN-VC: 
Non-parallel many-to-many voice conversion 
\\with star generative adversarial networks
}
\name{
Hirokazu Kameoka, Takuhiro Kaneko, Kou Tanaka, Nobukatsu Hojo
}
\address{NTT Communication Science Laboratories, NTT Corporation, Japan}
\begin{document}
%
\maketitle
\begin{abstract}
This paper proposes a method that allows non-parallel many-to-many voice conversion (VC) by using a variant of a generative adversarial network (GAN) called StarGAN. 
Our method, which we call StarGAN-VC, is noteworthy in that it 
(1) requires no parallel utterances, transcriptions, or time alignment procedures for speech generator training, (2) simultaneously learns many-to-many mappings across different attribute domains using a single generator network, (3) is able to generate converted speech signals quickly enough to allow real-time implementations and (4) requires only several minutes of training examples to generate reasonably realistic-sounding speech.
Subjective evaluation experiments on a non-parallel many-to-many speaker identity conversion task
revealed that the proposed method obtained higher sound quality and speaker similarity than a state-of-the-art method based on variational autoencoding GANs.
\end{abstract}
\section{Introduction}
\label{sec:intro}

Voice conversion (VC) is a technique for converting 
para/non-linguistic information
contained in a given utterance while preserving linguistic information.
This technique can be applied to various
tasks such as speaker-identity modification for text-to-speech
(TTS) systems \cite{Kain1998}, speaking assistance \cite{Kain2007,Nakamura2012}, speech enhancement \cite{Inanoglu2009,Turk2010,Toda2012}, and pronunciation conversion \cite{Kaneko2017c}.

One successful VC framework involves statistical methods
based on Gaussian mixture models (GMMs) \cite{Stylianou1998,Toda2007,Helander2010}.  
Recently, a neural network (NN)-based framework based on feed-forward deep NNs \cite{Desai2010,Mohammadi2014,YSaito2017bshort}, recurrent NNs \cite{Sun2015}, and generative adversarial nets (GANs) \cite{Kaneko2017c}, and an exemplar-based framework based on non-negative matrix factorization (NMF) \cite{Takashima2013,Wu2014} have also proved successful. 
Many conventional VC methods including those mentioned above require accurately aligned parallel 
source and target speech data. However, in many scenarios, 
it is not always possible to collect parallel utterances. 
Even if we could collect such data, we typically need to perform 
time alignment procedures, which becomes relatively
difficult when there is a large acoustic gap between
the source and target speech. Since many frameworks are weak as regards the misalignment
found with parallel data, careful pre-screening and
manual correction may be required to make these frameworks work reliably.
To bypass these restrictions, this paper is concerned with developing a non-parallel  
VC method, which requires no parallel utterances, transcriptions, or time alignment procedures.

In general, the quality and conversion effect obtained with non-parallel methods
are usually limited compared with methods using parallel data
due to the disadvantage related to the training condition. 
Thus, developing non-parallel methods with as high an audio quality and conversion effect as parallel methods can be very challenging. 
Recently, some attempts have been made to develop non-parallel methods \cite{Chen2014,Nakashika2014a,Nakashika2014b,Nakashika2015short,Blaauw2016,Hsu2016,Hsu2017,Xie2016,Kinnunen2017,Kaneko2017d,vandenOord2017bshort,Hashimoto2017short,YSaito2018bshort}.
For example, a method using automatic speech recognition (ASR) was proposed in \cite{Xie2016}.
The idea is to convert input speech under the restriction that the posterior state probability of the acoustic model of an ASR system is preserved so that the transcription of the converted speech becomes consistent with that of the input speech. 
Since the performance of this method depends heavily on the quality of the acoustic model of ASR,
it can fail to work if ASR does not function reliably. 
A method using i-vectors \cite{Dehak2011}, known as a feature for speaker verification, was recently proposed in \cite{Kinnunen2017}.
Conceptually, the idea is to
shift the acoustic features of input speech towards target speech in the i-vector space so that
the converted speech is likely to be recognized as the target speaker by a speaker recognizer.
While this method is also free from parallel data, one limitation is that 
it is applicable only to speaker identity conversion tasks.

Recently, a framework based on conditional variational autoencoders (CVAEs) \cite{Kingma2014a,Kingma2014b} was proposed in \cite{Hsu2016,YSaito2018bshort}.
As the name implies, variational autoencoders (VAEs) are a probabilistic counterpart of autoencoders (AEs), consisting of encoder and decoder networks.
CVAEs \cite{Kingma2014b} are an extended version of VAEs where the encoder and decoder networks can take an auxiliary variable $c$ as an additional input. By using acoustic features as the training examples and the associated attribute labels as $c$, 
the networks learn how to convert an attribute of source speech to a target attribute according to the attribute label fed into the decoder. 
This CVAE-based VC approach is notable in that it is completely free from parallel data and works even with unaligned corpora. 
However, one well-known problem as regards VAEs is that outputs from the decoder tend to be oversmoothed. 
For VC applications, this can be problematic since it usually results in poor quality buzzy-sounding speech. 

One powerful framework that can potentially overcome the weakness of VAEs involves GANs \cite{Goodfellow2014short}.
GANs offer a general framework for training a generator network in such a way that it can 
deceive a real/fake discriminator network. 
While they have been found to be effective for use with image generation, 
in recent years they 
have also been employed with notable success for various speech processing tasks \cite{Kaneko2017a,YSaito2018a,Pascual2017short,Kaneko2017b,Kaneko2017c,Oyamada2018}.
We previously reported a non-parallel VC method using a GAN variant called cycle-consistent GAN (CycleGAN) \cite{Kaneko2017d}, which
was originally proposed as a method for translating images using unpaired training examples \cite{Zhu2017,Kim2017,Yi2017}. 
This method, which we call CycleGAN-VC, is designed to
learn the mapping $G$ of acoustic features from one attribute $X$ to another attribute $Y$, 
its inverse mapping $F$, and 
a discriminator $D$, 
whose role is to distinguish the acoustic features of converted speech from those of real speech, 
through a training loss 
combining an adversarial loss and a cycle consistency loss.
Although this method was shown to work reasonably well, 
one major limitation is that it only learns one-to-one mappings. 
With a lot of VC application scenarios, it is desirable to obtain many-to-many mappings. 
One naive way of applying CycleGAN to many-to-many VC tasks would be to train different $G$ and $F$ pairs for all pairs of attribute domains. 
However, this may be ineffective since 
all attribute domains are common in the sense that they represent speech and so there must be common latent features that can be shared across different domains. 
In practice, the number of parameters will increase quadratically with the number of attribute domains, making parameter training challenging particularly when there are a limited number of training examples in each domain. 

A common limitation of CVAE-VC and CycleGAN-VC is that at test time the attribute of the input speech must be known. As for CVAE-VC, the source attribute label $c$ must be fed into the encoder of the trained CVAE and with CycleGAN-VC, the source attribute domains at training and test times must be the same. 

To overcome the shortcomings and limitations of CVAE-VC \cite{Hsu2016} and CycleGAN-VC \cite{Kaneko2017d}, this paper 
proposes a non-parallel many-to-many VC method using
a recently proposed novel GAN variant called StarGAN \cite{Choi2017short}, which 
offers the advantages of CVAE-VC and CycleGAN-VC concurrently. 
Unlike CycleGAN-VC and as with CVAE-VC, our method, which we call StarGAN-VC, is 
capable of simultaneously learning many-to-many mappings using a single encoder-decoder type generator network $G$ where the attributes of the generator outputs are controlled by an auxiliary input $c$. 
Unlike CVAE-VC and as with CycleGAN-VC, StarGAN-VC uses an adversarial loss 
for generator training 
to encourage 
the generator outputs to become indistinguishable from real speech
and ensure that the mappings between each pair of attribute domains
will preserve linguistic information.
It is also noteworthy that unlike CVAE-VC and CycleGAN-VC, StarGAN-VC does not require any information about the attribute of 
the input speech at test time.

The VAE-GAN framework \cite{Larsen2015short} is perhaps another natural way of overcoming the weakness of VAEs.
A non-parallel VC method based on this framework has already been proposed in \cite{Hsu2017}.
With this approach, an adversarial loss derived using a GAN discriminator is incorporated into the training loss to encourage the decoder outputs of a CVAE to be indistinguishable from real speech features. 
Although the concept is similar to our StarGAN-VC approach, 
we will show in \refsec{experiments} that our approach outperforms this method 
in terms of both the audio quality and conversion effect.

Another related technique worth noting is 
the vector quantized VAE (VQ-VAE) approach \cite{vandenOord2017bshort}, 
which has performed impressively in non-parallel VC tasks. 
This approach 
is particularly notable in that it offers a novel way of overcoming the weakness of VAEs by using
the WaveNet model \cite{vandenOord2016short}, a sample-by-sample neural signal generator, to devise both the encoder and decoder of a discrete counterpart of CVAEs.
The original WaveNet model is a recursive model that makes it possible to predict the distribution of a sample conditioned on the samples the generator has produced. 
While a faster version \cite{vandenOord2017ashort} has recently been proposed,
it typically requires huge computational cost to generate a stream of samples, which can cause difficulties when implementing real-time systems.
The model is also known to require a huge number of training examples to be able to generate natural-sounding speech.
By contrast, our method is noteworthy in that it is able to generate signals quickly enough to allow real-time implementation and requires only several minutes of training examples to generate reasonably realistic-sounding speech.

The remainder of this paper is organized as follows.
We briefly review the formulation of CycleGAN-VC in \refsec{cyclegan-vc},
present the idea of StarGAN-VC in \refsec{stargan-vc} and 
show experimental results in \refsec{experiments}.

\section{CycleGAN Voice Conversion}
\label{sec:cyclegan-vc}

Since the present method is an extension of CycleGAN-VC, which we proposed previously \cite{Kaneko2017d},  
we start by briefly reviewing its formulation.

Let $\Vec{x}\in \mathbb{R}^{Q\times N}$ and $\Vec{y}\in\mathbb{R}^{Q\times M}$
be acoustic feature sequences of speech belonging to attribute domains $X$ and $Y$, respectively,
where $Q$ is the feature dimension and $N$ and $M$ are the lengths of the sequences.
The aim of CycleGAN-VC is to learn 
a mapping $G$ that converts the attribute of $\Vec{x}$ into $Y$
and a mapping $F$ that does the opposite. Now, we introduce discriminators $D_X$ and $D_Y$,
whose roles are to predict whether or not their inputs are the acoustic features of real speech belonging to $X$ and $Y$,
and define 
\begin{align}
\mathcal{L}_{\rm adv}^{D_Y}(D_Y) 
=&- 
\mathbb{E}_{\Vec{y}\sim p_Y(\Vec{y})}[\log D_Y(\Vec{y})]\nonumber\\
&-\mathbb{E}_{\Vec{x}\sim p_X(\Vec{x})}[\log (1- D_Y(G(\Vec{x})))],
\label{eq:cyclegan-advloss_dy}
\\
\mathcal{L}_{\rm adv}^{G}(G) 
=&\mathbb{E}_{\Vec{x}\sim p_X(\Vec{x})}[\log (1- D_Y(G(\Vec{x})))],
\label{eq:cyclegan-advloss_g}
\\
\mathcal{L}_{\rm adv}^{D_X}(D_X) 
=&-
\mathbb{E}_{\Vec{x}\sim p_X(\Vec{x})}[\log D_X(\Vec{x})]\nonumber\\
&-
\mathbb{E}_{\Vec{y}\sim p_Y(\Vec{y})}[\log (1- D_X(F(\Vec{y})))],
\label{eq:cyclegan-advloss_dx}
\\
\mathcal{L}_{\rm adv}^{F}(F) 
=&
\mathbb{E}_{\Vec{y}\sim p_Y(\Vec{y})}[\log (1- D_X(F(\Vec{y})))],
\label{eq:cyclegan-advloss_f}
\end{align}
as the adversarial losses for $D_Y$, $G$, $D_X$ and $F$, 
respectively.
$\mathcal{L}_{\rm adv}^{D_Y}(D_Y)$ 
and 
$\mathcal{L}_{\rm adv}^{D_X}(D_X) $
measure how indistinguishable 
$G(\Vec{x})$ and $F(\Vec{y})$ are 
from acoustic features of real speech belonging to $Y$ and $X$. 
Since the goal of $D_X$ and $D_Y$ is to correctly distinguish 
the converted feature sequences obtained via $G$ and $F$
from real speech feature sequences, 
$D_X$ and $D_Y$ attempt to minimize these losses to avoid being fooled by $G$ and $F$.
Conversely, since one of the goals of $G$ and $F$ is to generate realistic-sounding speech that is indistinguishable from real speech, $G$ and $F$ attempt to maximize these losses or minimize 
$\mathcal{L}_{\rm adv}^{G}(G)$ and $\mathcal{L}_{\rm adv}^{F}(F)$ to fool $D_Y$ and $D_X$.
It can be shown that the output distributions of $G$ and $F$ trained in this way will match the empirical distributions 
$p_Y(\Vec{y})$ and $p_X(\Vec{x})$. 
Note that since 
$\mathcal{L}_{\rm adv}^{G}(G)$ and $\mathcal{L}_{\rm adv}^{F}(F)$ 
are minimized when $D_Y(G(\Vec{x}))\simeq 1$ and $D_X(F(\Vec{y}))\simeq 1$,
we can also use 
$- \mathbb{E}_{\Vec{x}\sim p_X(\Vec{x})}[\log D_Y(G(\Vec{x}))]$
and 
$- \mathbb{E}_{\Vec{x}\sim p_X(\Vec{x})}[\log D_Y(G(\Vec{x}))]$ 
as the adversarial losses for $G$ and $F$.

As mentioned in \refsec{intro}, training $G$ and $F$ using only the adversarial losses does not guarantee that 
$G$ or $F$ will preserve the linguistic information of the input speech since there are infinitely many mappings 
that will induce the same output distributions. 
To further regularize these mappings, we introduce a cycle consistency loss 
\begin{align}
\mathcal{L}_{\rm cyc}(G,F) 
&= \mathbb{E}_{\Vec{x}\sim p_X(\Vec{x})}[\| F(G(\Vec{x})) - \Vec{x} \|_1]
\nonumber\\
&+\mathbb{E}_{\Vec{y}\sim p_Y(\Vec{y})}[\| G(F(\Vec{y})) - \Vec{y} \|_1],
\end{align}
to encourage $F(G(\Vec{x}))\simeq \Vec{x}$ and $G(F(\Vec{y}))\simeq \Vec{y}$. 
With the same motivation, we also consider an identity mapping loss
\begin{align}
\mathcal{L}_{\rm id}(G,F) 
&= \mathbb{E}_{\Vec{x}\sim p_X(\Vec{x})}[\| F(\Vec{x}) - \Vec{x} \|_1]
\nonumber\\
&+\mathbb{E}_{\Vec{y}\sim p_Y(\Vec{y})}[\| G(\Vec{y}) - \Vec{y} \|_1],
\end{align}
to ensure that inputs to $G$ and $F$ 
are kept unchanged when the inputs already belong to $Y$ and $X$.
The full objectives of CycleGAN-VC to be minimized with respect to 
$G$, $F$, $D_X$ and $D_Y$ 
are thus given as 
\begin{align}
\mathcal{I}_{G,F}(G,F) =&
\mathcal{L}_{\rm adv}^{G}(G) +
\mathcal{L}_{\rm adv}^{F}(F) 
\nonumber\\
&+ 
\lambda_{\rm cyc}
\mathcal{L}_{\rm cyc}(G,F) + 
\lambda_{\rm id}
\mathcal{L}_{\rm id}(G,F),\\
\mathcal{I}_{D}(D_X,D_Y) =&
\mathcal{L}_{\rm adv}^{D_X}(D_X)
+
\mathcal{L}_{\rm adv}^{D_Y}(D_Y),
\end{align}
where $\lambda_{\rm cyc}\ge 0$ and $\lambda_{\rm id}\ge 0$ are 
regularization parameters, which weigh the importance of 
the cycle consistency loss and the identity mapping loss relative to the adversarial losses.

\begin{figure*}[t!]
\centering
  \begin{minipage}{.52\linewidth}
  \centerline{\includegraphics[width=.98\linewidth]{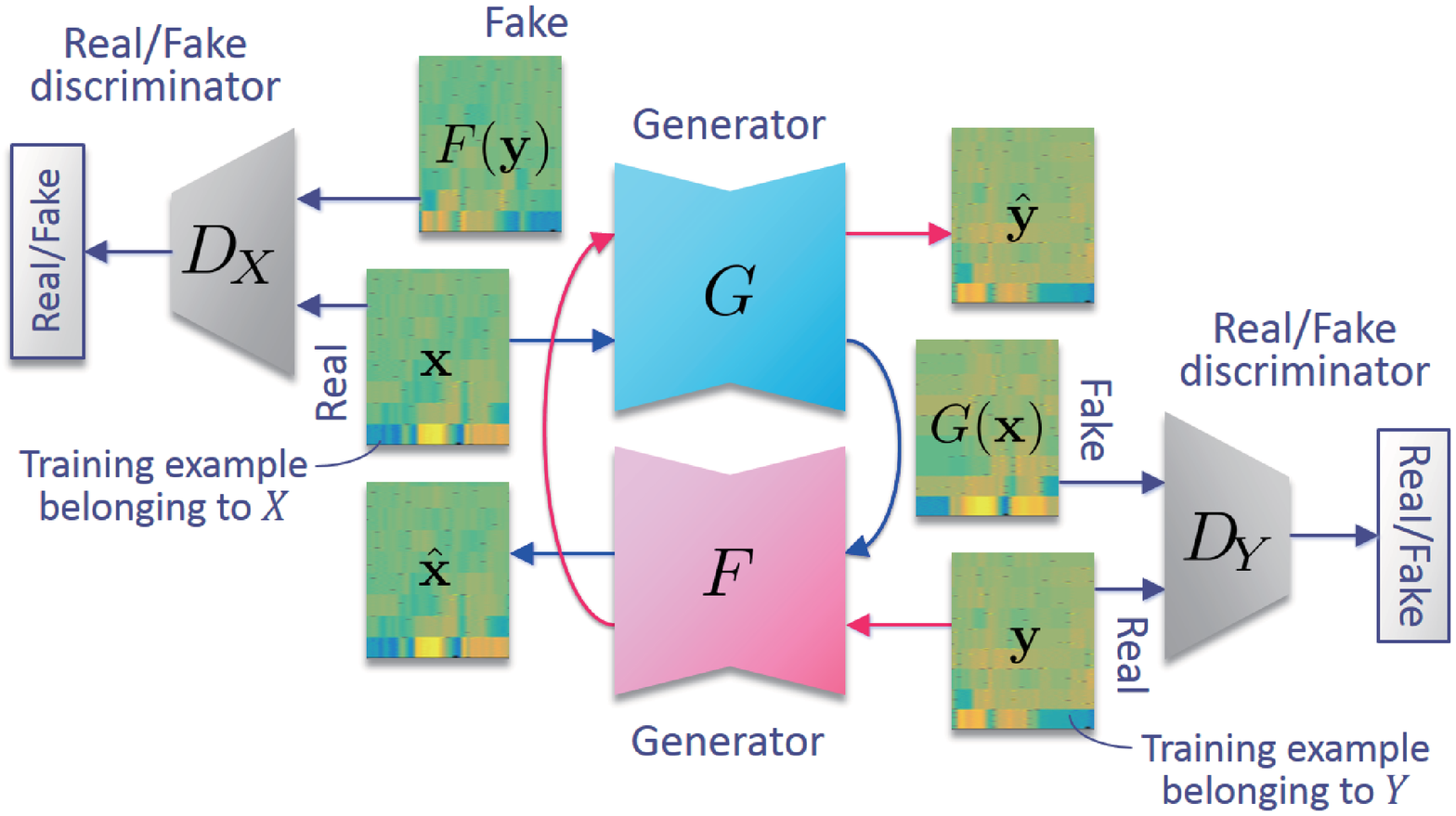}}
  \vspace{-1ex}
  \caption{Concept of CycleGAN training.}
  \label{fig:cyclegan-vc}
  \end{minipage}
  \begin{minipage}{.47\linewidth}
  \centerline{\includegraphics[width=.98\linewidth]{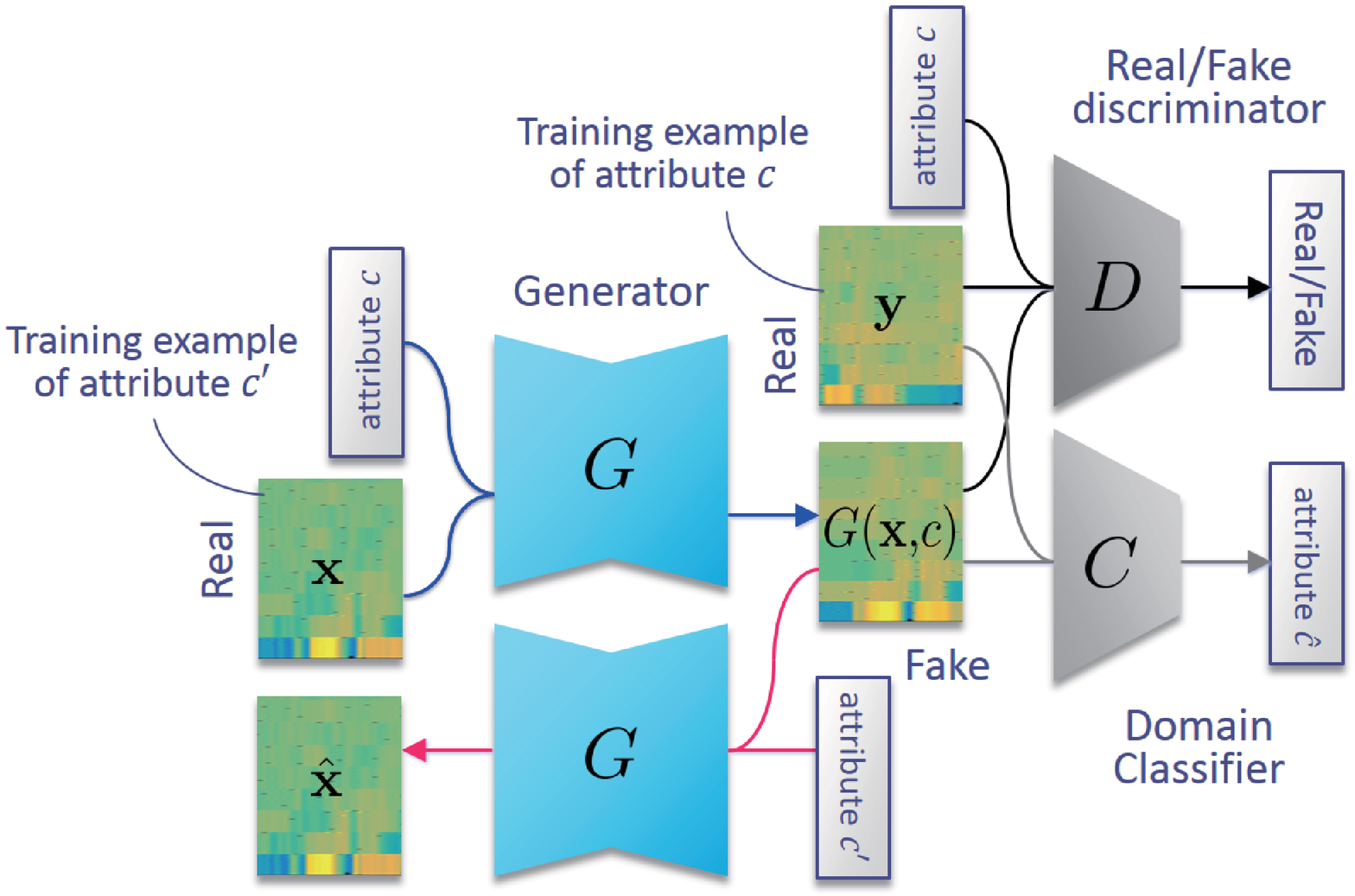}}
  \vspace{-1ex}
  \caption{Concept of StarGAN training.}
  \label{fig:stargan-vc}
  \end{minipage}
\end{figure*}

\section{StarGAN Voice Conversion}
\label{sec:stargan-vc}

While CycleGAN-VC allows the generation of natural-sounding speech when a sufficient number of training examples are available, one limitation is that it only learns one-to-one-mappings. 
Here, we propose using StarGAN \cite{Choi2017short} to develop a method that allows  
non-parallel many-to-many VC. We call the present method StarGAN-VC.

\subsection{Training objectives}

Let $G$ be a generator that takes an acoustic feature sequence $\Vec{x}\in \mathbb{R}^{Q\times N}$ with an arbitrary attribute and a target attribute label $c$ as the inputs and generates an acoustic feature sequence $\hat{\Vec{y}} = G(\Vec{x},c)$. We assume that a speech attribute comprises one or more categories, each consisting of multiple classes. We thus represent $c$ as a concatenation of one-hot vectors, each of which is filled with 1 at the index of a class in a certain category and with 0 everywhere else. 
For example, if we consider speaker identities as the only attribute category, 
$c$ will be represented as a single one-hot vector, where each element is associated with a different speaker.
One of the goals of StarGAN-VC is to make 
$\hat{\Vec{y}} = G(\Vec{x},c)$ as realistic as real speech features
and belong to attribute $c$. 
To realize this, we introduce a real/fake discriminator $D$ as with CycleGAN 
and a domain classifier $C$, 
whose role is to predict to which classes an input belongs. 
$D$ is designed to produce a probability $D(\Vec{y},c)$ that an input $\Vec{y}$ is a real speech feature whereas 
$C$ is designed to produce class probabilities $p_C(c|\Vec{y})$ of $\Vec{y}$.

\noindent
{\bf Adversarial Loss:}
First, we define
\begin{align}
\mathcal{L}_{\rm adv}^D(D) =& 
- \mathbb{E}_{c\sim p(c), \Vec{y}\sim p(\Vec{y}|c)}[\log D(\Vec{y},c)] 
\nonumber\\
&- \mathbb{E}_{\Vec{x}\sim p(\Vec{x}), c\sim p(c)}
[\log (1-D(G(\Vec{x},c),c))],
\label{eq:advloss_d}
\\
\mathcal{L}_{\rm adv}^G(G) =&
- \mathbb{E}_{\Vec{x}\sim p(\Vec{x}), c\sim p(c)}
[\log D(G(\Vec{x},c),c)],
\label{eq:advloss_g}
\end{align}
as adversarial losses for discriminator $D$ and generator $G$, respectively, 
where $\Vec{y}\sim p(\Vec{y}|c)$ denotes a training example of an acoustic feature sequence of real speech with attribute $c$ and 
$\Vec{x}\sim p(\Vec{x})$ denotes that with an arbitrary attribute. 
$\mathcal{L}_{\rm adv}^D(D)$ takes a small value when $D$ correctly classifies 
$G(\Vec{x},c)$ and $\Vec{y}$ as fake and real speech features 
whereas
$\mathcal{L}_{\rm adv}^G(G)$ takes a small value when $G$ successfully deceives $D$ so that 
$G(\Vec{x},c)$ is misclassified as real speech features by $D$.
Thus, we would like to minimize $\mathcal{L}_{\rm adv}^D(D)$ with respect to $D$ and
minimize $\mathcal{L}_{\rm adv}^G(G)$ with respect to $G$.

\noindent
{\bf Domain Classification Loss:} 
Next, we define
\begin{align}
\mathcal{L}_{\rm cls}^C(C) = &
- \mathbb{E}_{c\sim p(c), \Vec{y}\sim p(\Vec{y}|c)}
[\log p_C(c|\Vec{y})],
\\
\mathcal{L}_{\rm cls}^G(G) = &
- \mathbb{E}_{\Vec{x}\sim p(\Vec{x}), c\sim p(c)}
[\log p_C(c|G(\Vec{x},c))],
\end{align}
as domain classification losses for classifier $C$ and generator $G$.
$\mathcal{L}_{\rm cls}^C(C)$ and $\mathcal{L}_{\rm cls}^G(G)$ take small values
when $C$ correctly classifies $\Vec{y}\sim p(\Vec{y}|c)$ and 
$G(\Vec{x},c)$ as belonging to attribute $c$. 
Thus, we would like to minimize $\mathcal{L}_{\rm cls}^C(C)$ with respect to $C$ 
and $\mathcal{L}_{\rm cls}^G(G)$ with respect to $G$.

\noindent
{\bf Cycle Consistency Loss:} 
Training $G$, $D$ and $C$ using only the losses presented above does not guarantee that 
$G$ will preserve the linguistic information of input speech. 
To encourage $G(\Vec{x},c)$ to be a bijection, we introduce a cycle consistency loss to be minimized
\begin{multline}
\mathcal{L}_{\rm cyc}(G) \\
= 
\mathbb{E}_{c'\sim p(c), \Vec{x}\sim p(\Vec{x}|c'), c\sim p(c)}
[\| G(G(\Vec{x},c),c') - \Vec{x}\|_\rho],
\end{multline}
where 
$\Vec{x}\sim p(\Vec{x}|c')$ denotes a training example of an acoustic feature sequence of real speech with attribute $c'$ and $\rho$ is a positive constant.
We also consider an identity mapping loss
\begin{align}
\mathcal{L}_{\rm id}(G) = 
\mathbb{E}_{c'\sim p(c), \Vec{x}\sim p(\Vec{x}|c')}
[\|G(\Vec{x},c') - \Vec{x} \|_\rho],
\end{align}
to ensure that an input into $G$ will remain unchanged when the input already belongs to the target attribute $c'$.

To summarize, 
the full objectives of StarGAN-VC 
to be minimized with respect to $G$, $D$ and $C$
are given as 
\begin{align}
\mathcal{I}_G(G) =
&\mathcal{L}_{\rm adv}^{G}(G)
+
\lambda_{\rm cls}
\mathcal{L}_{\rm cls}^{G}(G)
\nonumber\\
&+
\lambda_{\rm cyc}
\mathcal{L}_{\rm cyc}(G)
+
\lambda_{\rm id}
\mathcal{L}_{\rm id}(G),
\\
\mathcal{I}_D(D) =
&\mathcal{L}_{\rm adv}^{D}(D),
\\
\mathcal{I}_C(C) =
&\mathcal{L}_{\rm cls}^{C}(C),
\end{align}
respectively, 
where $\lambda_{\rm cls}\ge 0$, $\lambda_{\rm cyc}\ge 0$ and $\lambda_{\rm id}\ge 0$ are 
regularization parameters, which weigh the importance of the domain classification loss,
the cycle consistency loss and the identity mapping loss relative to the adversarial losses.

\subsection{Conversion process}
\label{subsec:conversion}

As an acoustic feature vector, we use mel-cepstral coefficients computed 
from a spectral envelope obtained using WORLD \cite{Morise2016}. 
After training $G$, we can convert the acoustic feature sequence $\Vec{x}$ 
of an input utterance with
\begin{align}
\hat{\Vec{y}} = G(\Vec{x}, c),
\end{align}
where $c$ denotes the target attribute label.
A na\"ive way of obtaining a time-domain signal is simply to use $\hat{\Vec{y}}$ to reconstruct a signal with a vocoder.
Instead of directly using $\hat{\Vec{y}}$,
we can also use 
the reconstructed feature sequence 
\begin{align}
\hat{\Vec{y}}' = 
G(\Vec{x}, c'),
\end{align}
to obtain a time-domain signal
if the attribute $c'$ of the input speech is known.
By using $\hat{\Vec{y}}$ and $\hat{\Vec{y}}'$, 
we can obtain a sequence of spectral gain functions. 
Once we obtain the spectral gain functions, 
we can reconstruct a time-domain signal 
by multiplying 
the spectral envelope of input speech by
the spectral gain function frame-by-frame
and resynthesizing the signal using a vocoder. 

\subsection{Network architectures}

One of the key features of our approach including \cite{Kaneko2017c,Kaneko2017d}
 is that we consider a generator that 
takes an acoustic feature sequence instead of a single-frame acoustic feature as an input and 
outputs an acoustic feature sequence of the same length.
This allows us to 
obtain conversion rules that capture time dependencies.
While RNN-based architectures are a natural choice for modeling time series data, 
we use a convolutional neural network (CNN)-based architecture to design $G$ as detailed below.
The generator $G$ consists of encoder and decoder networks where 
only the decoder network takes an auxiliary input $c$.
We also design $D$ and $C$ to take acoustic feature sequences as inputs
and generate sequences of probabilities.

\noindent
{\bf Generator:}
Here, we treat an acoustic feature sequence $\Vec{x}$ as an image of size $Q \times N$ with 
$1$ channel and use 2D CNNs to construct $G$, 
as they are suitable for parallel computations. 
Specifically, we use a gated CNN \cite{Dauphin2017}, which 
was originally introduced to model word sequences for language modeling 
and was shown to outperform long short-term memory (LSTM) language models trained in a similar setting. 
We previously applied gated CNN architectures for voice conversion \cite{Kaneko2017c,Kaneko2017d} and audio source separation \cite{Li2018}, and their effectiveness has already been confirmed.
In the encoder part, 
the output of the
$l$-th hidden layer 
is described as a linear projection 
modulated by an output gate
\begin{align}
\Vec{h}_l &= (\W_l * \Vec{h}_{l-1} + \Vec{b}_l) \odot 
\sigma
(\V_l * \Vec{h}_{l-1} + \Vec{d}_l),
\label{eq:enc_glu}
\end{align}
where 
$\W_l \in \mathbb{R}^{D_l\!\times\! D_{l-1}\!\times\! Q_l\!\times\! N_{l}}$, 
$\Vec{b}_l\in \mathbb{R}^{D_l}$, 
$\V_l\in \mathbb{R}^{D_l\! \times\! D_{l-1}\!\times\! Q_l \!\times\! N_{l}}$ and 
$\Vec{d}_l\in \mathbb{R}^{D_l}$ are
the generator network parameters to be trained, and 
$\sigma$ denotes the elementwise sigmoid function.
Similar to LSTMs,
the output gate multiplies each element of 
$\W_l * \Vec{h}_{l-1} + \Vec{b}_l$
and control what information should be propagated through the hierarchy of layers.
This gating mechanism is called Gated Linear Units (GLU).
In the decoder part, 
the output of the $l$-th 
hidden layer is given by
\begin{align}
\Vec{h}_{l-1}' &= [\Vec{h}_{l-1};\Vec{c}_{l-1}],
\label{eq:dec_glu_1}
\\
\Vec{h}_l &= (\W_l * \Vec{h}_{l-1}' + \Vec{b}_l) \odot 
\sigma
(\V_l * \Vec{h}_{l-1}' + \Vec{d}_l),
\label{eq:dec_glu_2}
\end{align}
where
$[\Vec{h}_{l};\Vec{c}_{l}]$ means the concatenation of 
$\Vec{h}_{l}$ and $\Vec{c}_{l}$
along the channel dimension, and
$\Vec{c}_l$ is 
a 3D array consisting of a $Q_l$-by-$N_l$ tiling of copies of $c$ in the feature and time dimensions.
The input into the 1st layer of $G$ is $\Vec{h}_0 = \Vec{x}$ and
the output of the final layer is given as a regular linear projection
\begin{align}
\Vec{h}_{L-1}' &= [\Vec{h}_{L-1};\Vec{c}_{L-1}],\\
\hat{\Vec{y}} &= \W_L * \Vec{h}_{L-1}' + \Vec{b}_L.
\end{align}
It should be noted that the entire architecture is fully convolutional with no fully-connected layers, which allows us to take an entire sequence with an arbitrary length as an input and convert the entire sequence. 

\noindent
{\bf Real/Fake Discriminator:}
We leverage the idea of PatchGANs \cite{Isola2017} to
devise a real/fake discriminator $D$, which classifies 
whether local segments of an input feature sequence are real or fake. 
More specifically, we devise $D$ using a gated CNN, which takes 
an acoustic feature sequence $\Vec{y}$ and an attribute label $c$ as inputs and 
produces a sequence of probabilities that measures how likely each segment of $\Vec{y}$ is to be real speech features of attribute $c$.
The output of the $l$-th layer of $D$ is given in the same way as \refeqs{dec_glu_1}{dec_glu_2} and 
the final output $D(\Vec{y},c)$ is given by the product of all these probabilities. 
See \refsec{experiments} for more details.

\noindent
{\bf Domain Classifier:} 
We also devise a domain classifier $C$ using a gated CNN, which takes 
an acoustic feature sequence $\Vec{y}$ and 
produces a sequence of class probability distributions that measures how likely each segment of $\Vec{y}$ is to belong to attribute $c$.
The output of the $l$-th layer of $C$ is given in the same way as \refeq{enc_glu} and the final output $p_C(c|\Vec{y})$ is given by the product of all these distributions.
See \refsec{experiments} for more details.


\begin{figure*}[t!]
\centering
  \centerline{\includegraphics[width=.9\linewidth]{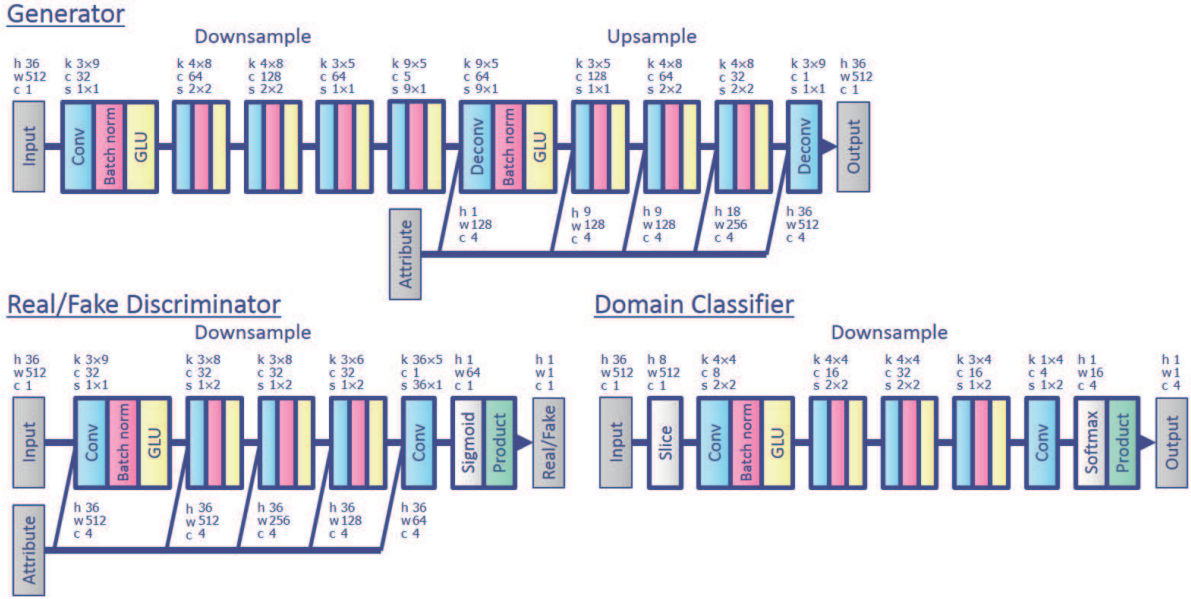}}
  \vspace{-1ex}
  \caption{Network architectures of generator $G$, real/fake discriminator $D$ and domain classifier $C$. Here, the inputs and outputs of $G$, $D$ and $C$ are interpreted as images, where ``h'', ``w'' and ``c'' denote the height, width and channel number, respectively. ``Conv'', ``Batch norm'', ``GLU'', ``Deconv'' ``Sigmoid'', ``Softmax'' and ``Product'' denote convolution, batch normalization, gated linear unit, transposed convolution, sigmoid, softmax, and product pooling layers, respectively. ``k'', ``c'' and ``s'' denote the kernel size, output channel number and stride size of a convolution layer, respectively. Note that all the networks are fully convolutional with no fully connected layers, thus allowing inputs to have arbitrary sizes.}
\label{fig:netarch}
\end{figure*}

\section{Subjective evaluation}
\label{sec:experiments}

\begin{figure*}[t!]
\centering
\begin{minipage}{.75\linewidth}
  \centerline{\includegraphics[width=.98\linewidth]{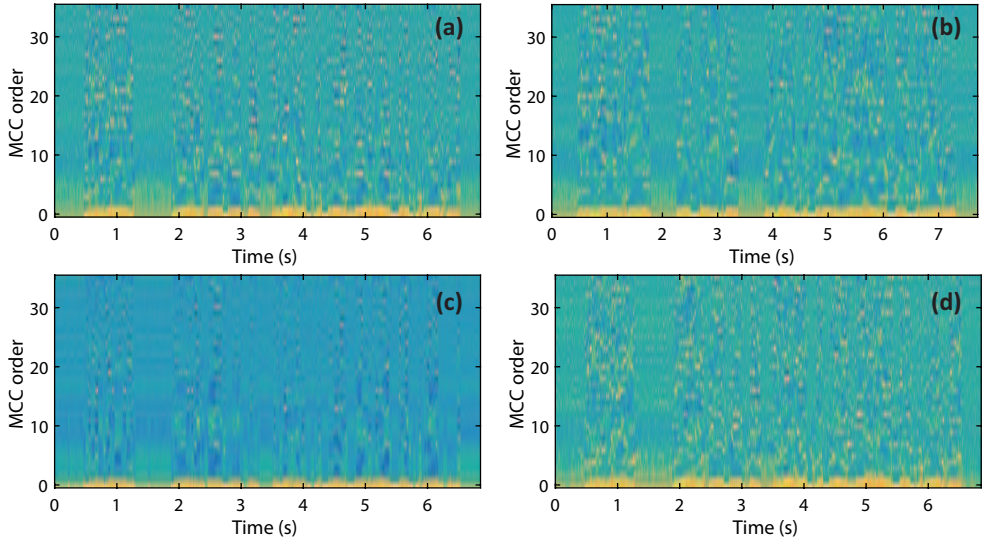}}
  \vspace{-2ex}
  \caption{Examples of acoustic feature sequences of (a) source speech, (c) converted speech obtained with  the baseline method and (d) converted speech obtained with the proposed method, along with an acoustic feature sequence of (b) the target speaker uttering the same sentence.}
\label{fig:mcep}
\end{minipage}
\end{figure*}

\begin{figure}[t!]
\centering
\begin{minipage}{.75\linewidth}
  \centerline{\includegraphics[height=6.2cm]{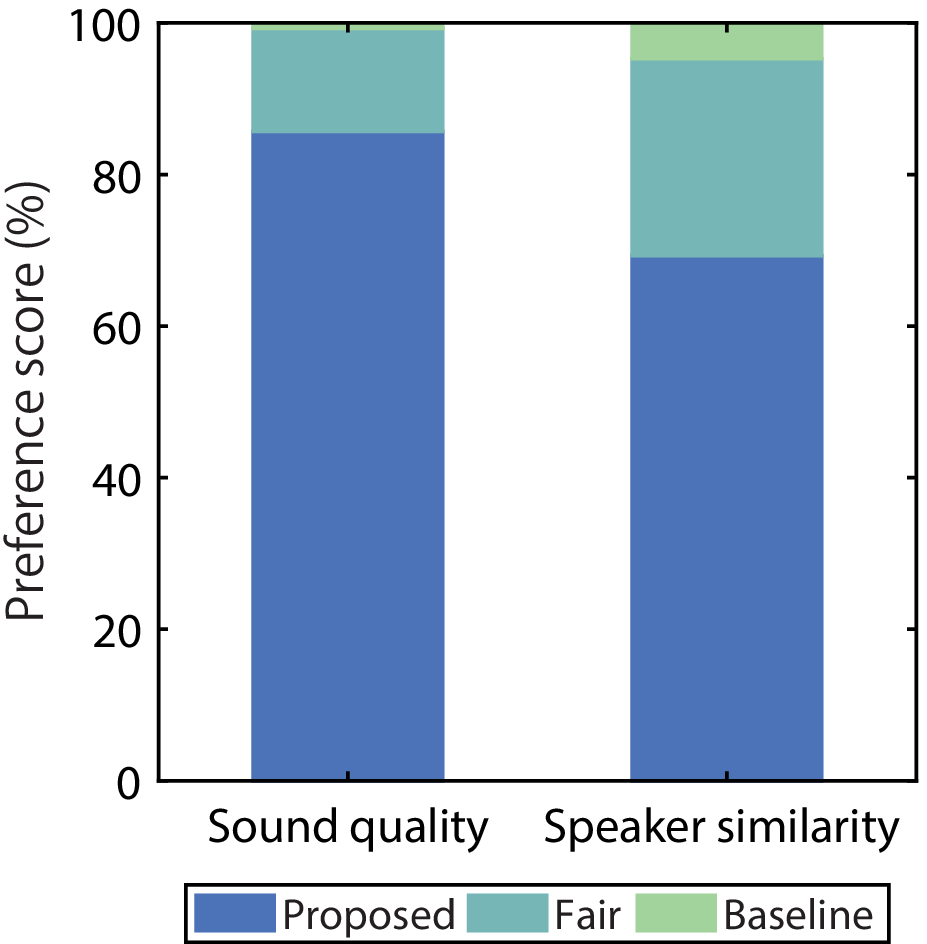}}
\caption{Results of the AB test for sound quality and the ABX test for speaker similarity.}
\label{fig:abx}
\end{minipage}
\end{figure}

To confirm the performance of the proposed method, 
we conducted subjective evaluation experiments on a non-parallel many-to-many speaker identity conversion task. 
We used the Voice Conversion Challenge (VCC) 2018 dataset \cite{Lorenzo-Trueba2018short},
which consists of recordings of six female and six male US English speakers. 
We used a subset of speakers for training and evaluation. 
Specifically, we selected two female speakers, `VCC2SF1' and `VCC2SF2', 
and two male speakers, `VCC2SM1' and `VCC2SM2'. 
Thus, $c$ is represented as a four-dimensional one-hot vector and 
there were twelve different combinations of source and target speakers in total.
The audio files for each speaker were manually
segmented into 116 short sentences (about 7 minutes)
where 81 and 35 sentences (about 5 and 2 minutes) were provided
as training and evaluation sets, respectively. 
All the speech signals were sampled at 22050 Hz. 
For each utterance, a spectral envelope,
a logarithmic fundamental frequency (log $F_0$), and
aperiodicities (APs) were extracted every 5 ms using the
WORLD analyzer \cite{Morise2016}. 36 mel-cepstral coefficients (MCCs) were then extracted from 
each spectral envelope. 
The $F_0$ contours were converted using the logarithm Gaussian normalized
transformation described in \cite{Liu2007}. The aperiodicities were used directly without
modification. The network configuration is shown in detail in \reffig{netarch}.
The signals of the converted speech were obtained using the method described in \refsubsec{conversion}.

We chose the VAEGAN-based approach \cite{Hsu2017} as a comparison for our experiments. 
Although we would have liked to exactly replicate the implementation of this method, 
we made our own design choices owing to missing details of the network configuration and hyperparameters.
We conducted an AB test to compare the sound quality of the converted speech samples
and an ABX test to compare the similarity to target speaker of the converted speech samples, where ``A'' and ``B'' were converted speech samples obtained with the proposed and baseline methods and ``X'' was a real speech sample of a target speaker.
With these listening tests, 
``A'' and ``B'' were presented in random orders to eliminate bias in the order of stimuli. 
Eight listeners participated in our listening tests. 
For the AB test for sound quality, 
each listener was presented \{``A'',``B''\} $\times$ 20 utterances, 
and for the ABX test for speaker similarity,
each listener was presented \{``A'',``B'',``X''\} $\times$ 24 utterances.
Each listener was then asked to select ``A'', ``B'' or ``fair'' for each utterance.
The results are shown in \reffig{abx}.
As the results show, 
the proposed method significantly outperformed the baseline method in terms of 
both sound quality and speaker similarity.
\reffig{mcep} shows an example of the MCC sequences of 
source, reconstructed, and converted speech.
Audio samples are provided at \url{http://www.kecl.ntt.co.jp/people/kameoka.hirokazu/Demos/stargan-vc/}.


\section{Conclusion}
\label{sec:conclusion}

This paper proposed a method that allows non-parallel many-to-many VC 
by using a novel GAN variant called StarGAN. 
Our method, which we call StarGAN-VC, is noteworthy in that it 
(1) requires no parallel utterances, transcriptions, or time alignment procedures for speech generator training, (2) simultaneously learns many-to-many mappings across different voice attribute domains using a single generator network, (3) is able to generate signals of converted speech quickly enough to allow real-time implementations and (4) requires only several minutes of training examples to generate reasonably realistic-sounding speech.
Subjective evaluation experiments on a non-parallel many-to-many speaker identity conversion task
revealed that the proposed method obtained higher sound quality and speaker similarity than a baseline method based on a VAE-GAN approach.


\small
\bibliographystyle{IEEEbib}
\bibliography{Kameoka2018arXiv05}

\end{document}